\documentclass[preprint,showpacs,preprintnumbers,amsmath,amssymb]{revtex4-1}

\usepackage{color} 
\usepackage{graphicx} 

\usepackage{dcolumn}
\usepackage{bm}

  \newcommand{\nn}{\nonumber}
\begin{document}

\title{Instability diagram of the massive gauge quantum fields around the nonlinear massive classical wave solution}

\author{Yoshio Kitadono}
 \email{kitadono@ncut.edu.tw}
\affiliation{ 
	Graduate Institute of Precision Manufacturing and Liberal Education Center,
	National Chin-Yi University of Technology,\\
	No.~57, Sec.~II, Zhongshan Road, Pingling, Taiping District, Taichung, 411030, Taiwan, Republic of China}
\author{Tomohiro Inagaki}
\email{inagaki@hiroshima-u.ac.jp}
\affiliation{ 
	Information Media Center and Core Research for Energetic Universe,
	Hiroshima University,\\
	No.~1-3-2, Kagamiyama, Higashi-Hiroshima, Hiroshima, 739-8521, Japan}

\date{\today}

\begin{abstract}
 The stability and instability in the time dynamics of quantum fluctuation of the massive gauge field coupling to the nonlinear massive wave solution is studied. In particular, the instability in the transverse polarization and longitudinal polarization modes are obtained as the two dimensional plot of the initial field value parameter and the spatial momentum of the quantum massive gauge fields with the help of the theory of the Hill's equation. We present the formalism to analyze the massive gauge field by taking into account Proca constraint and we found that our formalism can predict the time dynamics of the unstable quantum mode by the Floquet index. 
 The resulting polarization-resolved Floquet maps show that the transverse modes possess only narrow parametric-resonance bands, whereas the longitudinal modes exhibit substantially broader regions generated by both parametric and spinodal instabilities. We also find additional low-momentum instability regions for the $W$ boson that are absent or strongly suppressed in the $Z$ sector.
 \end{abstract}

\keywords{Higgs potential, Classical solution, Massive gauge fields,  Hill's equation}
\maketitle

\section{Introduction \label{Sec1}}
The generation of masses of elementary particles is the important aspect of the standard model (SM) of elementary particles. In particular, the existence of very heavy particles as the mediator of the weak interaction can be understood by the mechanism of the generating masses in the standard model. The idea of the generation of the masses without violating the gauge symmetries, which is known as so-called "Higgs mechanism", was proposed in 1964 \cite{Englert.Brout.1964,Higgs.1964,Guralnik.Hagen.Kibble.1964} and the existence of the Higgs boson was predicted. In 2012, the ATLAS and CMS collaborations at the large hadron collider (LHC) announced the discovery of the Higgs boson with the mass $m_H \simeq 125$~GeV at the LHC \cite{Higgs.ATLAS.2012, Higgs.CMS.2012}. The Higgs mechanism in the SM predicts the existence of the Higgs boson with the spin-0 and parity-even. This prediction was also tested in the experiments \cite{Higgs.SP.ATLAS.2016, Higgs.SP.CMS.2016} and the prediction of the SM is consistent with the experimental data. Moreover, many experimental data in LHC seem to be consistent with the theoretical calculations based on the SM.

Although many extension of the SM have been proposed theoretically, we do not consider new models, but consider a new aspect of the Higgs potential in the SM in our article. To be more precise, we investigate the quantum theory of particles coupling to classical solutions in the Higgs potential. Such model independent study will be important as well as model study, because the study of a new aspect of the Higgs potential in the SM will lead a new knowledge to the electroweak phase transition and related topics in elementary particle physics. 

The classical solution in the Higgs potential to be discussed in our article is the elliptic solution described by the Jacobi's elliptic functions. If the field oscillation of the Higgs field is bounded near the minimum of the Higgs potential, then the field oscillation is described by trigonometric functions like sine and cosine. However, if the field oscillation is not bounded near the minimum of the potential in general, then the dynamical solution is described by the elliptic solution. The replacement of the trigonometric solution with the elliptic solution is important because we can keep the non-linearity of the Higgs potential in the classical solution. 

We can find out the earlier research of the elliptic solutions in literature as the study of classical solutions in the equation of motion (EOM) of the gauge theory \cite{Treat.1971}. For example, S.~Coleman suggested $\mbox{SU}(2)$ gauge theory can give a nonlinear wave solution \cite{Coleman.1979}, the nonlinear massive solution of the pure $\mbox{SU}(2)$ gauge theory was obtained in \cite{Corrigan.1977, Oh.1985, Baseyan.1979, Matinyan.1981} which is related to Jacobi's elliptic function, and the solution of the scalar theory was obtained in \cite{Actor.1979}, the new type of solution given by Weierstrass's function in \cite{Tsapalis.2016}. These work are purely theoretical investigations. On the other hand, for phenomenological purpose, the prediction of the Higgs mass based on non-linearity wave analysis was discussed \cite{Achilleos.2012}, the relation between classical nonlinear solution and quantum fields in the $\mbox{SU}(3)$ gauge theory was developed \cite{Frasca.2008.2009}, the quantum fluctuations around the nonlinear massive solution in the Higgs potential \cite{KI.PRD.2016} and the particle creation was discussed in \cite{KI.PLB.2024}. Beside, the idea of the generation of the electroweak scale based on the renormalization group and elliptic solutions were discussed in \cite{Frasca.2025.RGE, Frasca.2026.HYM}. 

Apart from earlier work of the elliptic solutions in EOM, more recently, the doubt of usefulness of the effective potential to describe "time dynamics" was raised, and its improvement by considering the effect of the particle creation of quantum fields and a new equilibrium state as the consequence of time dynamics were comprehensively studied in \cite{Herring.2024}. The essential idea is the effect of the particle creation in the effective potential to combine the time dynamics in the quantum field and the effective potential of the theory. As the result, the authors conjectured the existence of a new equilibrium state and some thermodynamic quantities, like entropy density and number density of scalar bosons.  Moreover, its extension to the case taking into account the thermal effect was discussed \cite{Herring.2025}. It will be interesting to apply this picture to study the electroweak phase transition, when the classical Higgs field starts to oscillate with the nonlinear massive wave. In \cite{KI.PRD.2025}, the authors studied the instability diagram of the quantum fluctuation of the Higgs field around the nonlinear massive solution with Jacobi's $dn$- and $cn$-type functions and particle creation. 

Nonadiabatic production of Standard Model bosons and fermions by a time-dependent Higgs like background has previously been investigated using $cn$-type elliptic solutions \cite{Casadio.2007}, while the polarization-dependent dynamics of the vector fields has been studied mainly in dark-sector settings \cite{Khan.2026,Dror.2010}. Here we instead consider the two $dn$-type branches, for which the Higgs field never vanishes and the $W$ and $Z$ bosons therefore remain massive throughout each oscillation. To our knowledge, the polarization-resolved Floquet structure of the physical electroweak gauge fields has not been systematically mapped for the non-vanishing nonlinear Higgs backgrounds. We construct these two-dimensional instability diagrams and find a clear polarization hierarchy: the transverse modes possess only narrow parametric-resonance bands, whereas the longitudinal modes exhibit substantially broader band- and horn-shaped regions containing both parametric and spinodal instabilities. The different electroweak couplings also generate distinct low-momentum instability structures for the $W$ and $Z$ bosons.

For simplicity, we will not take into account the effect of the expansion of the Universe, because we must first find out which parameter set triggers the instability of the system and we will consider the effect of the evolution of the Universe in the future. Such study will contribute to find out a new equilibrium state based on the new effective potential analysis taking into account the renormalization effect and the particle production effect discussed recently in Ref.~\cite{Herring.2024}. 

We will introduce the nonlinear massive wave solution as the classical solution of EOM in the Higgs potential in Sec.~\ref{Sec2}. In Sec.~\ref{Sec3}, we adapt the standard quantization formalism for a
massive gauge field with a time-dependent mass to the nonlinear Higgs background and derive the transverse and longitudinal mode equations, paying particular attention to the Proca constraint.
In Sec.~\ref{Sec4}, we derive the stability/instability diagram as the main result of the article with the help of mathematical theorem called Hill's equation theory. The instability diagram is given in the two dimensional plot as the function of the squared momentum of the quantum fluctuation and the initial field value parameter in the classical solution. We will discuss the important nature of the result in Sec.~\ref{Sec5}, namely, how to understand the instability of the quantum fluctuation with the theorem in the Floquet theorem in the Hill's equation. We summarize the article in \ref{Sec6} and point out some possible applications of our results.

\section{Nonlinear massive wave solution \label{Sec2}}
We consider the Higgs potential in the SM:
\begin{eqnarray}
\mathcal{L} &=&
 (D_{\mu}\Phi)^{*}(D^{\mu}\Phi) - V(\Phi^{*}\Phi) + \mathcal{L}_{\mathrm {kin}},\nn\\
  V(\Phi^{*}\Phi) &=& - \mu^2 \Phi^{*}\Phi
             + \lambda (\Phi^{*}\Phi)^2, \nn\\
 \mathcal{L}_{\mathrm{kin}} &=& -\frac{1}{4}F^{a}_{\mu\nu}F^{a\mu\nu}
 -\frac{1}{4}B_{\mu\nu}B^{\mu\nu},
\end{eqnarray}
where $D_{\mu}$ stands for the covariant derivative including $\mbox{SU}(2)_{\mathrm L}$ and $\mbox{U}(1)_{\mathrm Y}$ gauges fields, $V$ is the Higgs potential, $\mathcal{L}_{\mathrm{kin}}$ stands for the kinetic terms of the gauge fields, $F^{a\mu\nu}$ and $B^{\mu\nu}$ are the field strengths for each gauge group,
$\mu^2$ is the squared mass parameter and $\lambda$ is the self coupling of the Higgs field, and $\Phi$ is the $\mbox{SU}(2)_{\mathrm{L}}$ Higgs doublet in the scalar sector in the SM.
We do not consider the couplings between the scalar field to other fields classically, or alternatively we treat other fields as quantum fields except for the scalar field. 

The Higgs doublet $\Phi$ in the unitarity gauge can be expressed as
\begin{eqnarray}
	\Phi(x) \to 
\frac{1}{\sqrt{2}}\begin{pmatrix}
	0 \\
	\phi(x)
\end{pmatrix},
\end{eqnarray}
where $\phi(x)$ stands for the neutral component in the doublet. Then, as is well known, the Lagrangian reduces to
\begin{eqnarray}
 \mathcal{L} = \frac{1}{2}(\partial_{\mu}\phi)(\partial^{\mu}\phi) - V(\varphi) + \frac{g^2}{4}W^{+}_{\mu}W^{-\mu}\phi^2 
 + \frac{g^2+g^{\prime~2}}{8}Z_{\mu}Z^{\mu}\phi^2
 + \mathcal{L}_{\mathrm{kin}} 
 + \cdots, 
\end{eqnarray}
where the $g$ and  $g^{\prime}$ are the couplings for $\mbox{SU}(2)$ and $\mbox{U}(1)_{Y}$ gauge groups respectively, and
the Higgs potential $V(\phi)$ reduces to
\begin{eqnarray}
 V(\phi) = -\frac{\mu^2}{2}\phi^2 + \frac{\lambda}{4}\phi^4.
\end{eqnarray}
The above potential has the vacuum expectation value (VEV) at $v = \sqrt{\mu^2/\lambda}$ and the standard Higgs field $h(x)$ is introduced as the quantum fluctuation around VEV, $\phi(x)=v+h(x)$.

The EOM for pure scalar sector is given in 
\begin{eqnarray}
	\partial^2 \phi - \mu^2 \phi + \lambda \phi^3 = 0,
\end{eqnarray}
and the elliptic solution in EOM is obtained as ~\cite{KI.PRD.2016}:
\begin{eqnarray}
	\phi_{\mathrm{cl}}(x) 
= \left\{
\begin{matrix}
 \frac{\phi_0}{dn(p_1\cdot x, k_1)} \hspace{1cm}(0 < \tilde{\phi}_0 < 1),\\
 \phi_0 dn(p_2 \cdot x, k_2) \hspace{1cm}(1 < \tilde{\phi}_0 < \sqrt{2}), \\
 \phi_0 cn(p_3 \cdot x, k_3) \hspace{1cm}(\sqrt{2} < \tilde{\phi}_0),
\end{matrix}
\right. \label{eq.phicl.3cases}
\end{eqnarray} 
where $dn(z,k)$ and $cn(z,k)$ are the Jacobi's $dn$-type and $cn$-type elliptic functions with the variable $z$ and elliptic modulus $k$ \cite{math.dlmf, Abramowitz, Gradshteyn}, and $\phi_{0}=\tilde{\phi}_0v$ is the initial field value parameter, respectively. The explicit expressions of the modulus $k_{i}$ and the squared invariant mass $p^2_{i}$ for three oscillation regions, $i=1,2$ and $3$, are given as the function of $\phi_0$ in Ref.~\cite{KI.PLB.2024}.  The solution seems to be complicated and has less relation with the SM at first glance. However, it shows a tight connection to SM, namely, the solution $\phi_{\mathrm{cl}}$ naturally reduces to VEV $v$ for the limit $\phi_0 \to v$ \cite{KI.PRD.2016}. The physical interpretation is that this solution describes the propagation of the massive and nonlinear wave in spacetime as long as $\phi_0 \neq v$, and it is suitable to describe a quasi-excited state of the classical Higgs field.

In this article, we take the rest frame of the four momentum in the classical  solution, i.e., $p^{\mu}_{i} = (m_{\mathrm{cl}},\vec{0})$ for $i=1,2$ and $3$ with $m^2_{\mathrm{cl}}\equiv p^2_{i}$ for $i=1,2$ and $3$. Then the classical solution $\phi_{\mathrm{cl}}(x)$ reduces to the time dependent solution, $\phi_{\mathrm{cl}}(t)$. Actually, we can always take this Lorentz frame because the classical solution gives the massive dispersion relation, $p^2_{i} \neq 0$ for $i=1,2$ and $3$. Besides, although we do not explicitly consider the effect of the time evolution of the Universe, we assume that the effect of the nonlinear solution at the present Universe is sufficiently small enough due to the suppression of the scale factor of the Universe \cite{Turner.1983}.

\section{Basic formalism of quantization around the classical nonlinear massive solution \label{Sec3}}
We follow the formalism of the linear quantization in Ref.~\cite{Herring.2024} to discuss the stability/instability of the quantum fluctuation. We introduce the quantum fluctuation $h$ by
\begin{eqnarray}
	\phi(x) = \phi_{\mathrm{cl}}(x) + \hbar^{\frac{1}{2}}h(x),
\end{eqnarray}
and expand the Lagrangian density up to $\mathcal{O}(\hbar)$, we then obtain
\begin{eqnarray}
  \mathcal{L} 
&=& \frac{1}{2}(\partial_{\mu}\phi_{\mathrm cl})^2 - V(\phi_{\mathrm cl}) 
  + \hbar\left[ \frac{1}{2}(\partial h)^2 + \frac{\mu^2}{2}h^2 - \frac{3}{2}\lambda \phi^2_{\mathrm cl}h^2 \right]
  \nn\\
  &&
  + \frac{g^2}{4}W^{+}_{\mu}W^{-\mu}\phi^2_{\mathrm cl} + \frac{g^2+g^{\prime~2}}{8}\phi^2_{\mathrm cl}Z_{\mu}Z^{\mu} 
 + \mathcal{O}(\hbar^{\frac{3}{2}}), \label{eq.Lagrangian.linear}
\end{eqnarray}
where we count the charged weak gauge field $W$ and neutral weak gauge field $Z$ as $\mathcal{O}(\hbar^{1/2})$ contributions in the $\hbar$-expansion, hence $WW$ and $ZZ$ are counted as $\mathcal{O}(\hbar)$ contributions. The instability analysis for the scalar field $h$ was derived in Ref.~\cite{KI.PRD.2025} and we only consider the instability analysis for massive vector fields which couple to the classical Higgs field in the second line of Eq.~(\ref{eq.Lagrangian.linear}). We set $\hbar=1$ in our analysis hereafter. 

We can derive EOMs for the massive electroweak vector fields, $W$ and $Z$, by using Eq.~(\ref{eq.Lagrangian.linear}). The EOM for the massive charged $W$ boson field is given in
\begin{eqnarray}
 \partial^2 W^{-\mu}(x) - \partial^{\mu} (\partial \cdot W^{-}(x)) + g_W \phi^2_{\mathrm{cl}}(x) W^{-\mu}(x) = 0, \label{eq.EOM.W}
\end{eqnarray}
while the EOM for the massive neutral $Z$ boson field is given in
\begin{eqnarray}
  \partial^2 Z^{\mu}(x) - \partial^{\mu} (\partial \cdot Z(x)) + g_Z \phi^2_{\mathrm{cl}}(x) Z^{\mu}(x) = 0, \label{eq.EOM.Z}
\end{eqnarray}
where $g_W=g^2/4$ and $g_Z=(g^2+g^{\prime~2})/4$. 
Therefore, we consider the generalized EOM as the unified description for these two fields,
\begin{eqnarray}
 \partial^2 A^{\mu} - \partial^{\mu}\left(\partial \cdot A\right) + g_A \phi^2_{\mathrm{cl}}A^{\mu} = 0, \label{eq.massive.vector.EOM.ourA}
\end{eqnarray}
where $A^{\mu}$ stands for the massive vector bosons (either $Z$- or $W$-bosons) and $g_A$ stands for the coupling between the classical Higgs field to the massive vector field.

Before we discuss the EOM in our model, we briefly emphasize the special nature of massive vector field. Let $A^{\mu}$ be a massive vector field with mass $m$.  Then the typical EOM for the field is given in
\begin{eqnarray}
  \partial^2 A^{\mu} - \partial^{\mu}\left(\partial \cdot A\right) + m^2 A^{\mu} = 0. \label{eq.massive.vector.EOM.usualA}
\end{eqnarray}
Taking the divergence of the above equation, we obtain
\begin{eqnarray}
 \partial_{\mu} 
 \left[ 
\partial^2 A^{\mu} - \partial^{\mu}\left(\partial \cdot A\right) \right]
+ \partial_{\mu} (m^2 A^{\mu}) = 0,
\end{eqnarray}
where the first and second terms cancels out each other and we obtain $(\partial_{\mu}A^{\mu})=0$ for nonzero $m$. This condition, called Proca condition, is not the constraint on the choice of gauge, but it constrains the solution of EOM for the massive vector field $A^{\mu}$ for nonzero mass case. 

\subsection{Conventional quantization}
We briefly review the conventional approach of the quantization of the massive vector field by following the notation in Ref.~\cite{MaranonGonzalez.2023}. The complete set of the orthonormalized mode in Eq.~(\ref{eq.massive.vector.EOM.usualA})  for a given spatial momentum $\vec{k}$ can be obtained as
\begin{eqnarray}
 A^{r \mu}_{\vec{k}}(x) = \epsilon^{r \mu}(\vec{k})\frac{e^{-ik\cdot x}}{\sqrt{2(2\pi)^3\omega_k}}, \label{eq.func.form.Ak}
\end{eqnarray}
where $\omega_k = \sqrt{\vec{k}^2 + m^2}$ is the angular frequency, $\epsilon^{r\mu}(\vec{k})$ is the polarization vector with the polarization state $r$. We take $r=1,2$ as transverse polarizations, $r=3$ as the longitudinal polarization, and the transverse polarization vectors are given in
\begin{eqnarray}
 \epsilon^{r \mu}(\vec{k}) 
 = \left(0, \vec{\epsilon}^{~r}(\vec{k} \right) \hspace{1cm}(r=1,2), 
\end{eqnarray}
with $|\vec{\epsilon}^{~r}(\vec{k})|^2=1$, and $\vec{k}\cdot\vec{\epsilon}^{~r}(\vec{k})=0$ for $r=1,2$; while the longitudinal polarization vector is given in
\begin{eqnarray}
	\epsilon^{r \mu}(\vec{k}) = \left(\frac{k}{m}, \frac{\omega_k \vec{k}}{mk} \right) \hspace{1cm}(r=3),
\end{eqnarray}
with the notation $k\equiv|\vec{k}|$ and these polarization vectors are chosen to satisfy the relations, $k_{\mu}\epsilon^{r\mu}=0$ and $\epsilon^{r*}_{\mu}\epsilon^{s\mu}=-\delta^{rs}$. The normalization of the field in (\ref{eq.func.form.Ak}) is given by
\begin{eqnarray}
   -i\int d^3x \left( A^{*r\mu}_{\vec{k}}\dot{A}^{s}_{\vec{q}\mu}- \dot{A}^{*r\mu}_{\vec{k}}A^{s}_{\vec{q}\mu}, \right) \equiv \delta^{rs}\delta^3(\vec{k}-\vec{q}), \label{eq.normalization.A}
\end{eqnarray}
where $\dot{A}(t)=\partial_0A(t)$.

Integrating out momentum and summing polarization state, the quantization for the massive field $A^{\mu}$ is given in 
\begin{eqnarray}
 A^{\mu}(x) = \int \frac{d^3k}{\sqrt{2\omega_k(2\pi)^3}}\sum_{r=1}^{3}
 \left[ e^{-ik\cdot x}\epsilon^{r\mu}(\vec{k}) a^{r}_{\vec{k}}
+  e^{ik\cdot x}\epsilon^{*r\mu}(\vec{k}) a^{r\dagger}_{\vec{k}}
 \right],
\end{eqnarray}
where the normalization condition of the field gives the conventional commutation relation, $[a^{r}_{\vec{k}},a^{s\dagger}_{\vec{q}}]=\delta^{rs}\delta^3(\vec{k}-\vec{q})$, and other commutation relations are zero.

It is worthy noting that the mass term in the angular frequency $\omega_k=\sqrt{k^2+m^2}$ determined by Eq.~(\ref{eq.massive.vector.EOM.usualA}) is time independent in conventional method; while the corresponding "mass term" which will be determined by Eq.~(\ref{eq.massive.vector.EOM.ourA}) in our model seems to be time dependent because of the replacement, $m^2 \to g_A\phi^2_{\mathrm{cl}}$. 

\subsection{Quantization for transverse mode}
To discuss the quantization of the massive vector field which couples to the time-dependent nonlinear massive wave solution, all time dependence is encoded in the mode function. Typically, the quantization for the system coupling to time dependent classical field is used in the quantum field theory in expanding universe \cite{MukhanovText,BirrellText, ParkerText}.

In particular, to discuss our approach of the quantization of the massive vector field which couples to the time dependent nonlinear wave field, we split the quantization method to two ways, depending on the polarization modes. We follow the formalism in Ref.~\cite{MaranonGonzalez.2023} to manage the subtlety of the longitudinal mode. We begin the discussion with the definition of the orthonormalized mode for the massive vector field by 
\begin{eqnarray}
 A^{\mu r}_{\vec{k}} 
 \equiv
 \frac{e^{i\vec{k}\cdot \vec{x}}}{\sqrt{2(2\pi)^3}}
 \epsilon^{\mu r}(\vec{k},t)f^{r}_{k}(t), \label{eq.func.form.A}
\end{eqnarray}
where we assume the polarization vector $\epsilon^{\mu r}(\vec{k},t)$ and the mode function $f^{r}_k(t)$ which reflect all time dynamics of the system are given in
\begin{eqnarray}
 \epsilon^{\mu r}(\vec{k},t) 
 &=&
 \begin{pmatrix}
 	0, & \vec{\epsilon}^{~r}(\vec{k}) 
 \end{pmatrix}
\hspace{1cm}
f^{r}_k(t) = h_{k}(t),
\end{eqnarray}
for the transverse polarization mode $(r=1,2)$ and
\begin{eqnarray}
\epsilon^{\mu r}(\vec{k},t) 
&=&
 \begin{pmatrix}
 	\frac{kW(t)}{m\omega_k(t)} & \frac{\omega_k(t)\vec{k}}{mk} 
 \end{pmatrix}
 \hspace{1cm}
f^{r}_k(t) = \ell_{k}(t),
\end{eqnarray}
for the longitudinal polarization mode $(r=3)$, respectively. The time dependent angular frequency is defined by $\omega_k(t)=\sqrt{k^2+g_A\phi^2_{\mathrm{cl}}(t)}$, and $W(t)$ is the time-dependent auxiliary function which is determined later by the EOM and Proca constraint \cite{MaranonGonzalez.2023}. It is notable that the time dependence not only appears in the mode function $\ell_k(t)$, but also appears in the time component in the longitudinal polarization vector $\epsilon^{\mu r=3}$. 

Firstly, we derive the mode equation $h_k(t)$ for the transverse mode and its constraint. We substitute the expression in (\ref{eq.func.form.A}) into the normalization condition in (\ref{eq.normalization.A}), then we can derive the normalization condition for $f^{r=1,2}_k(t)=h_k(t)$,
\begin{eqnarray}
\dot{h}^{*}_{k}h_k
- h^{*}_k \dot{h}_k = 2i, \label{eq.Wronskian.hk}
\end{eqnarray}
where alternatively, this is called "Wronskian condition" for the mode function $h_k$. 

Next we go back to EOM in (\ref{eq.massive.vector.EOM.ourA}) to determine the time dynamics of the mode function $h_k(t)$. Using the expression of the divergence of the vector field, $(\partial_{\mu} A^{\mu})$, in ~(\ref{eq.func.form.A}) for $r=1,2$, 
\begin{eqnarray}
 \partial_{\nu} A^{\nu r}
= \frac{e^{i\vec{k}\cdot \vec{x}}}
       {\sqrt{2(2\pi)^3}}
\left[    
  \epsilon^{0r}(\vec{k})\dot{h}_k 
- i\vec{k}\cdot \vec{\epsilon}^{~r}(\vec{k})h_{k}
\right].
\end{eqnarray} 
and the relations for the transverse polarization, $\epsilon^{0r}=0$ and $\vec{k}\cdot \vec{\epsilon}^{~r}=0$ for $r=1,2$, then we conclude $\partial_{\nu} A^{\nu r} = 0$ for transverse mode. Then EOM reduces to Klein-Gordon type equation,
\begin{eqnarray}
 \partial^2 A^{\mu r} + g_A \phi^2_{\mathrm{cl}} A^{\mu r} = 0.
\end{eqnarray}
This equation can be converted into the differential equation for transverse mode function $h_k$:
\begin{eqnarray}
 \ddot{h}_k(t) + a_{k,T}(t)h_k(t)=0, \label{eq.modeeq.h}
\end{eqnarray}
with $a_{k,T}(t) = \omega^2_k(t)$. 

Finally we take into account the Proca constraint for the transverse mode. Taking the divergence of EOM in (\ref{eq.massive.vector.EOM.ourA}), we obtain the Proca constraint
\begin{eqnarray}
 \partial_{\mu}\left(g_A\phi^2_{\rm cl}(t)A^{\mu~r=1,2}\right)=0. \label{eq.Proca.constraint}
\end{eqnarray}
Expanding the above constraint, we have
\begin{eqnarray}
 2\phi_{\rm cl}(t)(\partial_{\mu}\phi_{\rm cl}(t))A^{\mu~r=1,2} + \phi^2_{\rm cl}(t)(\partial_{\mu}A^{\mu~r=1,2})=0.
\end{eqnarray}
Because $A^{0r}\propto \epsilon^{0r}=0$ and $\vec{\nabla}\phi_{\rm  cl}(t)\propto \vec{p} = 0$ for $i=1,2$ and $3$ in the rest frame of the four momentum of the classical Higgs field, we can conclude that the Proca constraint is automatically satisfied for the transverse mode. 

We can compare the above result for the transverse mode with the result for the scalar case discussed in \cite{KI.PRD.2025}. Actually, the structure of the differential equation for the mode function in the transverse mode is exactly same with the that for the scalar fluctuation. Hence we can use the same analysis based on the theory in Hill's equation in \cite{KI.PRD.2025} in analyzing the dynamics of the mode function $h_k(t)$.

\subsection{Quantization for longitudinal mode}
Secondly, we derive quantization of the mode function $\ell_k(t)$ for the longitudinal mode and its constraint by using the time-dependent-auxiliary field to manage the longitudinal mode which is introduced in \cite{MaranonGonzalez.2023}. We substitute the expression in (\ref{eq.func.form.Ak}) for the longitudinal mode into the normalization condition in (\ref{eq.normalization.A}), after some calculations, then the normalization relation between the mode function $\ell_k$ and polarization vector $\epsilon^{r\mu}$ is given in
\begin{eqnarray}
1 = \mbox{Im}
\left[ \epsilon^{\mu*r}\dot{\epsilon}^{r}_{\mu}
|\ell_k|^2
+ \epsilon^{\mu* r}\epsilon^{r}_{\mu} \ell^{*}_k \dot{\ell}_k
\right], \label{eq.Wronskian.r3}
\end{eqnarray}
where this result is consistent with the result in \cite{MaranonGonzalez.2023} when we set $a(t)=1$, because we do not consider the evolution of the universe and thus we can set the scale factor $a(t)$ to be unity in our analysis; while $a(t)\neq 1$ for the analysis in \cite{MaranonGonzalez.2023}.
 
To simplify the above result, we calculate $\epsilon^{\mu* r}\epsilon^{r}_{\mu}$ and $\epsilon^{\mu* r}\dot{\epsilon}^{r}_{\mu}$ and the results reduce to:
\begin{eqnarray}
 \epsilon^{r}_{\mu}\epsilon^{\mu r*} &=& 
   \frac{k^2|W|^2}{m^2\omega^2_k(t)}
   -\frac{\omega^2_k(t)}{m^2} \equiv -g_W(t),\nn\\
 \dot{\epsilon}^{r}_{\mu}\epsilon^{\mu r*} 
 &=& \frac{k^2}{m^2 \omega^2_k(t)}
 \left[W^{*}\dot{W}-|W|^2\frac{\dot{\omega}_k(t)}{\omega_k(t)}\right] - \frac{\omega_k(t)\dot{\omega}_k(t)}{m^2}, \label{eq.epsilon.contract.r3}
\end{eqnarray}
for $r=3$ case.
 Using the above results, the normalization condition is expressed by $W$ and $\ell_k$,
\begin{eqnarray}
 1 = \mbox{Im}
 \left[\frac{k^2}{m^2\omega^2_k(t)}|\ell_{k}|^2W^{*}\dot{W}-g_{W}(t)\ell^{*}_k\dot{\ell}_k \right], \label{eq.normalization.TL}
\end{eqnarray}
where we dropped the terms which have no imaginary terms.

Next we consider the Proca constraint for the longitudinal mode in (\ref{eq.Proca.constraint}). We rewrite the Proca constraint $\partial_{\mu} (\phi^2_{\rm cl}A^{\mu~r=3})=0$ in terms of $W(t)$. Expanding the constraint in (\ref{eq.Proca.constraint})
and using the rest frame of the four momentum in the nonlinear massive wave solution, we obtain
\begin{eqnarray}
 0 = 2\frac{\dot{\phi}_{\rm cl}}{\phi_{\rm cl}}A^{0r} + \partial_{\mu}A^{\mu r},
\end{eqnarray}
where we divided the result by $\phi^2_{\mathrm{cl}}(t)$. It is notable that $\phi_{\mathrm{cl}} \neq 0$ is guaranteed by the mathematical nature of the Jacobi's $dn$ function, while the Jacobi's $cn$ function periodically reduces to zero \cite{math.dlmf}. Hence, to avoid the possibility of $\phi_{\mathrm{cl}} = 0$, we constrain the initial value parameter $\phi_0$ in our analysis so that the relation $\phi_{\mathrm{cl}}\neq 0$ is realized. 

The second term in the above equation can be evaluated as
\begin{eqnarray}
 \partial_{\mu}A^{\mu r} 
 =
 A^{0r}\left[ \frac{\dot{\epsilon}^{r0}}{\epsilon^{r0}} + \frac{\dot{\ell}_k}{\ell_k}
 +\frac{ik^i\epsilon^{ri}}{\epsilon^{r0}}
 \right],
\end{eqnarray}
and combining all, the Proca constraint reduces to
\begin{eqnarray}
 0 = 2
 \frac{\dot{\phi}_{\rm cl}}{\phi_{\rm cl}} 
 + \frac{\dot{\epsilon}^{0r}}{\epsilon^{0r}} + \frac{\dot{\ell}_k}{\ell_k}
 +\frac{ik^i\epsilon^{ir}}{\epsilon^{0r}}.
\end{eqnarray}
Substituting 
$\epsilon^{0r}=kW/(m\omega_k)$ and $\epsilon^{ir}=\omega_kk^i/(mk)$ for $r=3$, the above constraint reduces to
\begin{eqnarray}
 0 = \left(2\frac{\dot{\phi_{\rm cl}}}{\phi_{\rm cl}} - \frac{\dot{\omega}_k}{\omega_k} + \frac{\dot{\ell}_k}{\ell_k}\right)W + \dot{W} + i\omega^2_k,
\end{eqnarray}
where the above equation constrains the relation between the auxiliary field $W(t)$ and its time derivative $\dot{W}(t)$.
Next we multiply $W^{*}$ to the above constraint and take the imaginary part, we obtain
\begin{eqnarray}
 \mbox{Im}(W^*\dot{W}) &=& 
 - \mbox{Im}\left(\frac{\dot{\ell}_k}{\ell_k}\right)|W|^2 
 - \omega^2_k\mbox{Re}(W).
\end{eqnarray}
Then we can eliminate $W\dot{W}$ term in the normalization condition in Eq.(\ref{eq.normalization.TL}) and we obtain 
\begin{eqnarray}
 -1 = \frac{k^2|\ell_k|^2}{m^2}\mbox{Re}(W)
 + \frac{\omega^2_k}{m^2}
 \mbox{Im}(\ell^*_k\dot{\ell}_k),
\end{eqnarray}
where the above equation determines how the time-dependence in the time component of the longitudinal polarization vector, $W(t)$, is related to the longitudinal mode function, $\ell_{k}(t)$, in our model. 

Finally we analyze EOM in (\ref{eq.massive.vector.EOM.ourA}) to derive the differential equation for $\ell_{k}(t)$. Taking $\mu=0$ component, then we have
\begin{eqnarray}
0 &=& 
-\partial^2_iA^{0r}
-\partial^0\partial_iA^{ir}
+ g_A\phi^2_{\rm cl}A^{0r},
\end{eqnarray}
and replacing $\partial_i$ with $ik^{i}$, then finally we obtain the relation between $A^{0r}$ and $A^{ir}$,
\begin{eqnarray}
A^{0r} = -\frac{i}{\omega^2_k}k^i\dot{A}^{r}_i, \label{eq.A0.Ai.relation}
\end{eqnarray}
where $r=3$. Eq.~(\ref{eq.A0.Ai.relation}) shows that the temporal component $A^0$ is nondynamical and is algebraically determined by the longitudinal spatial mode. Substituting this relation into the spatial components
of the field equation is equivalent to integrating out $A^0$ from the quadratic action \cite{Dror.2010}.

Rewriting the above relation to $W$ and $\ell_{k}$, we have
\begin{eqnarray}
 W = i\left(\frac{\dot{\omega}_k}{\omega_k}
 + \frac{\dot{\ell}_k}{\ell_k}
 \right),
\end{eqnarray}
alternatively taking real and imaginary part of $W$, the above equation can be rewritten as, 
\begin{eqnarray}
 \mbox{Re}(W) = -\mbox{Im}\left(\frac{\dot{\ell}_k}{\ell_k}\right), \hspace{1.5cm}
 \mbox{Im}(W) = \frac{\dot{\omega}_k}{\omega_k} + \mbox{Re}\left( \frac{\dot{\ell}_k}{\ell_k}\right).
\end{eqnarray}
To derive the equation for the mode function $\ell_k$, we set $\mu=i=1,2$ and $3$ components in (\ref{eq.massive.vector.EOM.ourA}), we obtain
\begin{eqnarray}
0=
 -\ddot{A}_i - \omega^2_kA_i
 +k^ik^j\frac{\ddot{A}_{j}}{\omega^2_k}
 -2k^ik^j\frac{\dot{A}_j\dot{\omega}_k}{\omega^3_k}
 +k^ik^jA_j.
\end{eqnarray}
Using the expressions of the time derivatives for $A^{r}_i$ for $r=3$, finally we obtain
\begin{eqnarray}
	0 = \ddot{\ell}_k 
	+ 2\frac{\dot{\phi}_{\rm cl}}{\phi_{\rm cl}}\dot{\ell}_k
	+\left[ g_A\left(\frac{\dot{\phi}^2_{\rm cl} +\phi_{\rm cl}\ddot{\phi}_{\rm cl}}{\omega^2_k}-g_A\frac{\phi^2_{\rm cl}\dot{\phi}^2_{\rm cl}}{\omega^4_k}\right)
	+\omega^2_k + 2k^2\frac{g_A\dot{\phi}^2_{\rm cl}}{\omega^4_{k}}\right]\ell_k.
\end{eqnarray}
To use the Floquet theorem in Hill's equation theory, we should eliminate $\dot{\ell}_k$ term with keeping $\ddot{\ell}_k$ term,
we perform the transformation called the standardization of the Hill's equation \cite{Hill.eq.book.1, Hill.eq.review.1}. In our case, we change the field variable, $\ell_k(t) \to L_{k}(t)$, through the relation,
\begin{eqnarray*}
 \ell_k(t) = \phi^n_{\rm cl}(t)L_{k}(t),
\end{eqnarray*}
with an integer $n$, then
after some calculation, we obtain
\begin{eqnarray}
 0 &=& \ddot{L}_k + 2(n+1)\frac{\dot{\phi}_{\rm cl}}{\phi_{\rm cl}}\dot{L}_k \nn\\ 
 &{}&+ \left[  n\frac{\ddot{\phi}_{\rm cl}}{\phi_{\rm cl}} 
         + n(n+1)\frac{\dot{\phi}^2_{\rm cl}}{\phi^2_{\rm cl}}
         + g_A\left(\frac{\dot{\phi}^2_{\rm cl} +\phi_{\rm cl}\ddot{\phi}_{\rm cl}}{\omega^2_k}-g_A\frac{\phi^2_{\rm cl}\dot{\phi}^2_{\rm cl}}{\omega^4_k}\right)
         +\omega^2_k + 2k^2\frac{g_A\dot{\phi}^2_{\rm cl}}{\omega^4_{k}}
         \right]L_k.
\end{eqnarray}
Therefore we can conclude that the choice $n=-1$ can eliminate the term of $\dot{L}_k$ from the differential equation and we obtain the Hill's-type equation: 
\begin{eqnarray}
 0 = \ddot{L}_k  
 + \left[  -\frac{\ddot{\phi}_{\rm cl}}{\phi_{\rm cl}} 
 + g_A\left(\frac{\dot{\phi}^2_{\rm cl} +\phi_{\rm cl}\ddot{\phi}_{\rm cl}}{\omega^2_k}-g_A\frac{\phi^2_{\rm cl}\dot{\phi}^2_{\rm cl}}{\omega^4_k}\right)
 +\omega^2_k + 2k^2\frac{g_A\dot{\phi}^2_{\rm cl}}{\omega^4_{k}}
 \right]L_k,
\end{eqnarray}
where this equation corresponds to Eq.(3.40) in \cite{MaranonGonzalez.2023}.
Alternatively, thanks to EOM for the classical field, $\ddot{\phi}_{\rm cl} 
  = \phi_{\rm cl}(\mu^2-\lambda\phi^2_{\rm cl})$,
the mode equation for $L_{k}(t)$ reduces to another form:
\begin{eqnarray}
0 &=& \ddot{L}_{k}(t) + a_{k,L}(t)L_{k}(t), \label{eq.modeeq.L}
\end{eqnarray}
with 
\begin{eqnarray}
a_{k,L}(t) 
&=& 
 -\lambda (v^2 -\phi^2_{\rm cl}) 
+ g_A\left(\frac{\dot{\phi}^2_{\rm cl} +\lambda \phi^2_{\rm cl}(v^2-\phi^2_{\rm cl})}{\omega^2_k}-g_A\frac{\phi^2_{\rm cl}\dot{\phi}^2_{\rm cl}}{\omega^4_k}\right)
+\omega^2_k + 2k^2\frac{g_A\dot{\phi}^2_{\rm cl}}{\omega^4_{k}}, \label{eq.modeeq.L.aLK}
\end{eqnarray}
where the mode equation for the longitudinal mode reflects the dynamics of the Higgs field through the term, $-\lambda(v^2-\phi^2)$. 

Finally we briefly mention the mass parameter $m$ in the longitudinal polarization vector $\epsilon^{\mu r=3}$. Using the relation $dn^2(z,k)=1-k^2sn^2(z,k)$ and the Fourier expansion of $sn^2(z,k)$ \cite{math.dlmf}:
\begin{eqnarray}
	sn^2(z,k) 
	= \frac{1}{k^2}\left( 1-\frac{E(k)}{K(k)} \right) 
	- \frac{2\pi^2}{k^2K^2(k)}\sum_{n=1}^{\infty} \frac{nq^{n}}{1-q^{2n}}\cos(2n\zeta),
\end{eqnarray}
with $q=\exp(-\pi K(k^{\prime})/K(k))$ and $k^{\prime}=\sqrt{1-k^2}$, $\zeta=\pi z/(2K(k))$, and $E(k)$ representing the complete elliptic integral of the second kind, we can define $m$ as the coupling $g_A$ times zero mode of the Fourier expansion of $dn^2$ function,
\begin{eqnarray}
	 m^2 = g_A \phi^2_0 \frac{E(k)}{K(k)},
\end{eqnarray}
for $dn$-type oscillation. We can define the mass parameter for $1/dn$-type oscillation by the same method. In our case, the elliptic modulus $k$ is given as the function of $\phi_0$ \cite{KI.PLB.2024} and massive vector boson mass $M_Z$ is recovered for the limit $\phi_0 \to v$ with $g_A=g_Z$ case.

\section{Instability diagram for quantum fluctuation \label{Sec4}}

\subsection{Hill's equation and Floquet theorem}
To analyze the stability and instability of solutions in Eqs.~(\ref{eq.modeeq.h}) and (\ref{eq.modeeq.L}), we use the Floquet theorem in Hill's equation. Thus we briefly review the essence of the theorem. In general, the Hill's equation is described by the differential equation,
\begin{eqnarray}
 y''(t) + a(t)y(t) = 0, \label{eq.Hill}
\end{eqnarray}
where $y''(t)$ stands for the second derivative of $y(t)$ with respect to $t$, $a(t)$ is a periodic function satisfying $a(t+T)=a(t)$ with a period $T$. The Floquet theorem gives the important behavior of the solution in the equation \cite{Hill.eq.book.1,Hill.eq.review.1}. According to the theorem, the solutions of the Hill's equation satisfy the translation relation,
\begin{eqnarray}
 y(t+T) = \rho y(t),
\end{eqnarray}  
with a complex constant $\rho$. Using the $n$-times translation, $y(t+nT)=\rho^ny(t)$, the behavior of $y(t)$ can be characterized by the magnitude of $|\rho|$. If $|\rho|>1$, then the solution is unstable, if $|\rho|<1$, then the solution is stable, and if $|\rho|=1$, then the solution reduces to either periodic or anti-periodic. The Floquet theorem gives the second order characteristic equation for $\rho$,
\begin{eqnarray}
 \rho^2 - \left( u_1(T) + u'_2(T) \right)\rho + 1 = 0, \label{eq.character.rho}
\end{eqnarray}
where $u_1(t)$ and $u_2(t)$ are the normalized solution defined by the initial conditions:
\begin{eqnarray}
 u_1(0) &=& 1, \hspace{1cm} u'_1(0) = 0, \nn\\
 u_2(0) &=& 0, \hspace{1cm} u'_2(0) = 1, \label{eq.u1.u2}
\end{eqnarray}
where these satisfy the Wronskian condition, $u_1{u}^{\prime}_2-{u}^{\prime}_1 u_2=1$.
Therefore, by analyzing the magnitude of the discriminant of the characteristic equation, $d=|u_1(T)+u'_2(T)|$, we can predict the time evolution of solutions of the Hill's equation. Moreover, the Floquet theorem gives that the two independent solutions of Eq.~(\ref{eq.Hill}), $y_1$ and $y_2$, are related to the solutions of the characteristic equations in Eq.~(\ref{eq.character.rho}),
$\rho_1$ and $\rho_2$, through the relation;
\begin{eqnarray}
 y_1(t) = e^{+i\nu t}\Pi_1(t), \hspace{1cm}
 y_2(t) = e^{-i\nu t}\Pi_2(t),
\end{eqnarray}
with the Floquet index $\nu$ and periodic functions $\Pi_1$ and $\Pi_2$. As a consequence, if $d>2$, then $\rho_1=e^{-i\nu T}$ and $\rho_2=e^{+i\nu T}$ reduce to real numbers and hence the solution shows a unstable behavior; while if $d<2$, then $\rho_1$ and $\rho_2$ reduce to complex numbers and hence the solution shows a stable behavior. The Floquet index can be evaluated by the relation $2\cos(\nu T)=\rho_1 + \rho_2$. 

Note that the Bloch theorem for a periodic potential problem in condensed matter physics \cite{Bloch.1929} is one of application of Floquet theorem to quantum physics, however, the typical solution in condensed matter physics is described by the stable solution \cite{KittelText}, while we can study the both stable and unstable solutions in this article.

\subsection{Instability of massive vector boson fields}
The mode equation for the transverse mode in (\ref{eq.modeeq.h}) can be rewritten by the dimensionless variables,
\begin{eqnarray}
	\frac{d^2}{d\tau^2}\tilde{h}_k(\tau)
	+ \tilde{a}_{k,T}(\tau)\tilde{h}_{k}(\tau) = 0, \label{eq.dimless.modeeq.h}
\end{eqnarray}
where we defined the dimensionless quantities;
\begin{eqnarray}
	\tau=m_{\rm cl}t, \hspace{1cm} \tilde{h}_k = \frac{h_k}{m^{-\frac{1}{2}}_{\rm cl}}, 
	\hspace{1cm}
	\tilde{a}_{k,T} = \frac{a_{k,T}}{m^2_{\mathrm{cl}}}. 
\end{eqnarray}
The expression of the coefficient $\tilde{a}_{k,T}$ is given in
\begin{eqnarray}
	\tilde{a}_{k,T}
	&=&\frac{k^2+g_A\phi^2_{\rm cl}}{m^2_{\rm cl}}
	\equiv  
		\tilde{k}^2 + \frac{g_Z\phi^2_{\rm cl}}{m^2_{\rm cl}},
\end{eqnarray} 
with $\tilde{k}^2=k^2/m^2_{\rm cl}$.
To analyze the stability/instability of solutions in Eq.~(\ref{eq.dimless.modeeq.h}), we check the period of $\tilde{a}_{k,T}(\tau)$. Using the fact that $\phi^2_{\mathrm{cl}}$ in $\tilde{a}_{k,T}$ is proportional to $dn^2(\tau,k)$ or $1/dn^2(\tau,k)$ in our analysis, and the mathematical nature of the Jacobi's elliptic functions $dn^2(z,k)$ \cite{Abramowitz,Gradshteyn,math.dlmf}, 
\begin{eqnarray}
 dn^2(z+2K(k),k) = dn^2(z,k),
\end{eqnarray}
where $K(k)$ stands for the complete elliptic integral of the first kind with the modulus $k$, we can derive the period $T$ of $\tilde{a}_{k,T}(\tau)$ as:
\begin{eqnarray}
 T = 
 \left\{
 \begin{matrix}
 	 2K(k_1) & (0<\tilde{\phi}_0<1), \\
 	 2K(k_2) & (1<\tilde{\phi}_0<\sqrt{2}),
 \end{matrix}
 \right.
\end{eqnarray}
where the numerical plot of the modulus $k_1$ and $k_2$ as the function of $\tilde{\phi}_0$ are given in \cite{KI.PRD.2025}. 
  
\begin{figure}[htbp]
\centering
	\begin{minipage}[b]{0.48\columnwidth}
    \centering	
	\includegraphics[width=\linewidth]{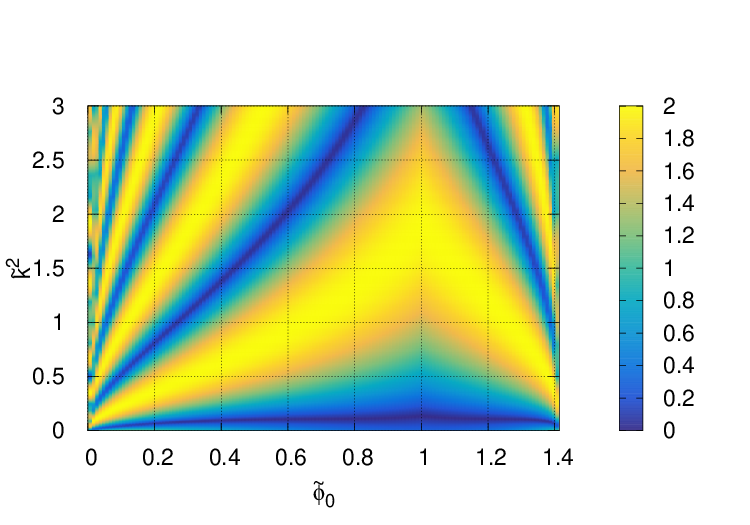}\\
		(a) Region with $d<2$
	\end{minipage}
	\begin{minipage}[b]{0.48\columnwidth}
	\centering
	\includegraphics[width=\linewidth]{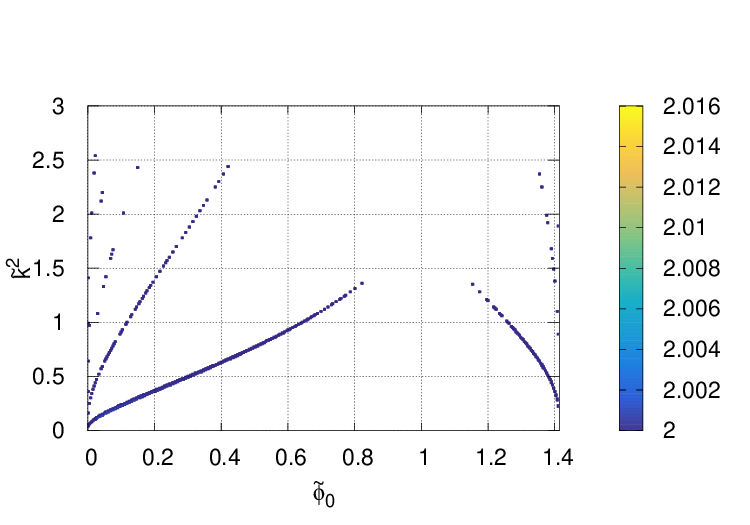}\\
		(b) Region with $d>2$
	\end{minipage}
	\caption{The plot of the absolute value of the discriminant $d$ for the transverse mode in Z field. \label{fig.disc.transverse.Z} }
\end{figure}

To plot the absolute value of the discriminant of the Hill's equation for transverse mode, we take the coupling $g_A$ as $g_A=g_Z=0.1374$ which corresponds to the coupling between $\phi_{\mathrm{cl}}$ and $ZZ$ fields, and this value reproduces the Z-boson mass \cite{PDG}. We take $\lambda=0.13$ which reproduces the Higgs mass measured at LHC \cite{Higgs.mass.new.ATLAS.2025, Higgs.mass.new.CMS.2025}. 
The numerical results are shown in (a) and (b) of Fig.~\ref{fig.disc.transverse.Z}. The numbers of divisions for $\tilde{\phi}_0$ and $\tilde{k}^2$ directions are $141$-points and $400$-points for (a), and $400$-points and $400$-points for (b), respectively.   
The result in (a) of Fig.~\ref{fig.disc.transverse.Z} shows that the most of parameters in $(\tilde{\phi}_0, \tilde{k}^2)$ plane give stable solutions in Eq.~(\ref{eq.dimless.modeeq.h}), while the result in (b) of Fig.~\ref{fig.disc.transverse.Z} shows the exception, i.e., some parameters which have line-shape structure give unstable solutions in Eq.~(\ref{eq.dimless.modeeq.h}). 

Similarly, the mode equation for the longitudinal mode can be written as
\begin{eqnarray}
	0 =  \frac{d^2}{d\tau^2}\tilde{L}_{k}(\tau)
	+ \tilde{a}_{k,L}(\tau)\tilde{L}_{k}(\tau), \label{eq.dimless.modeeq.L}
\end{eqnarray}
where the dimensionless mode function $\tilde{L}_k$ and the dimensionless coefficient $\tilde{a}_{k,L}$ are defined in
\begin{eqnarray}
	\tilde{L}_{k}(\tau) \equiv \frac{L_{k}(\tau)}{m^{\frac{1}{2}}_{\rm cl}},	
	\hspace{1cm}
	\tilde{a}_{k,L}(\tau) \equiv \frac{a_{k,L}(\tau)}{m^2_{\rm cl}}, 
\end{eqnarray}
where $\tilde{a}_{k,L}$ is given in
\begin{eqnarray}
	\tilde{a}_{k,L} = \frac{1}{\tilde{m}^2_{\rm cl}}
	\left[ 
	- \lambda(1-\tilde{\phi}^2_{\rm cl})
	+ g_{A}\left( \frac{(\frac{d\tilde{\phi}_{\rm cl}}{d\tau})^2}{\tilde{a}_{k,T}}
	+ \frac{\lambda\tilde{\phi}^2_{\rm cl}(1-\tilde{\phi_{\rm cl}^2})}{\tilde{m}^2_{\rm cl}\tilde{a}_{k,T}}
	- g_A \frac{\tilde{\phi}^2_{\rm cl}(\frac{d\tilde{\phi}_{\rm cl}}{d\tau})^2}{\tilde{m}^2_{\rm cl}\tilde{a}^2_{k,T}}
	\right)
	+ \tilde{m}^2_{\rm cl}\tilde{a}_{k,T} + 2g_A\frac{\tilde{k}^2(\frac{d\tilde{\phi}_{\rm cl}}{d\tau})^2}{\tilde{a}^2_{k,T}}
	\right], \label{eq.dimless.a.L}\nn\\
\end{eqnarray}
with $\tilde{m}^2_{\mathrm{cl}}=m^2_{\mathrm{cl}}/v^2$.

\begin{figure}[htbp]
	\centering
	\begin{minipage}[b]{0.48\columnwidth}
		\centering	
	\includegraphics[width=\linewidth]{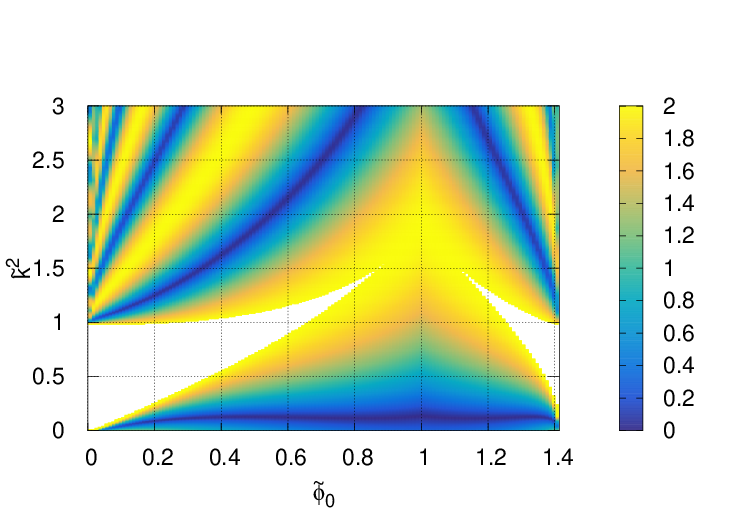}\\
		(a) Region with $d<2$ 
	\end{minipage}
	\begin{minipage}[b]{0.48\columnwidth}
		\centering
	\includegraphics[width=\linewidth]{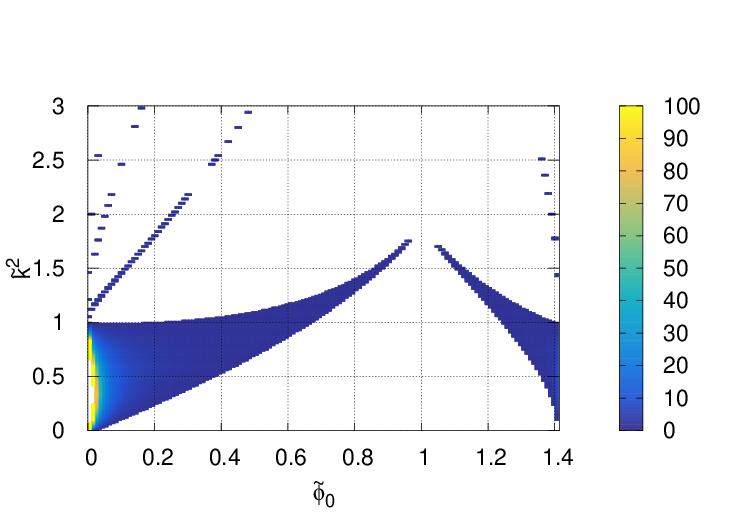}\\
		(b) Region with $d>2$
	\end{minipage}
	\caption{The plot of the absolute value of the discriminant $d$ for the longitudinal mode in Z boson. \label{fig.disc.longitudinal.Z}}
\end{figure}

We plot the absolute value of the discriminant of the Hill's equation for longitudinal mode in (a) and (b) of Figs.~\ref{fig.disc.longitudinal.Z}, where the numbers of divisions for the parameter spaces are same with that of Fig.~\ref{fig.disc.transverse.Z}.
The results in (a) and (b) of Fig.~\ref{fig.disc.longitudinal.Z} show stable and unstable parameter regions in $(\tilde{\phi}_0, \tilde{k}^2)$ space. In particular, we can easily see that the instability region in (b) of Fig.~\ref{fig.disc.longitudinal.Z} is broader than that in (b) of Fig.~\ref{fig.disc.transverse.Z}. As a consequence, the solution in (\ref{eq.dimless.modeeq.L}) tends to be more unstable than that in (\ref{eq.dimless.modeeq.h}), as we will discuss the detail in the next section. Note that the numerical behaviors of $d$ at $\tilde{\phi}_0 \approx 0$ with $0<\tilde{k}^2 <1.0$ in (b) of Fig.~\ref{fig.disc.longitudinal.Z} are extremely singular and thus we eliminated the region with the value $d>10^2$.

\subsection{Solutions of mode equations}
Because the mode function $h_k$ and its derivative $\dot{h}_k$ have the mass dimension $-1/2$ and $1/2$ respectively, and thus we consider the initial conditions for the dimensionless solutions in (\ref{eq.dimless.modeeq.h}),
\begin{eqnarray}
	\tilde{h}_k(0) = \frac{1}{\sqrt{\tilde{\omega}_{k}(0)}}, 
	\hspace{1cm}
	\tilde{h}^{\prime}_k(0) = -i\sqrt{\tilde{\omega}_{k}(0)}.
\end{eqnarray}
where $\tilde{h}^{'}_k$ stands for the derivative of $\tilde{h}_k$, and $\tilde{\omega}_{k}(0)=\sqrt{\tilde{a}_{k,T}(0)}$. This choice is the analogy of the adiabatic mode function to discuss the particle production in \cite{Herring.2024}. 

The general solution with the above initial conditions can be constructed by the linearly independent solutions in (\ref{eq.dimless.modeeq.h});
\begin{eqnarray}
	\tilde{h}_k(\tau) = \tilde{h}_k(0)u_1(\tau) + \tilde{h}^{\prime}_k(0)u_2(\tau),
\end{eqnarray}
where $u_1$ and $u_2$ are defined in (\ref{eq.u1.u2}).

To study the stability/instability of for $h_k(\tau)$, we take two set of parameters:
\begin{eqnarray}
	&(\mbox{i})&~\mbox{stable~point:}~(\tilde{\phi}_0,\tilde{k}^2)=(1.4,0.38), \nn\\
	&(\mbox{ii})&~\mbox{unstable~point:}~(\tilde{\phi}_0, \tilde{k}^2)=(1.4,0.37),
\end{eqnarray}
where these parameters give $d=1.998$ for the parameter (i) and $d=2.001$ (Floquet index=$0.004463i$) for the parameter (ii) respectively.

\begin{figure}[htbp]
		\centering
	\begin{minipage}[b]{0.48\columnwidth}
		\centering	
	\includegraphics[width=\linewidth]{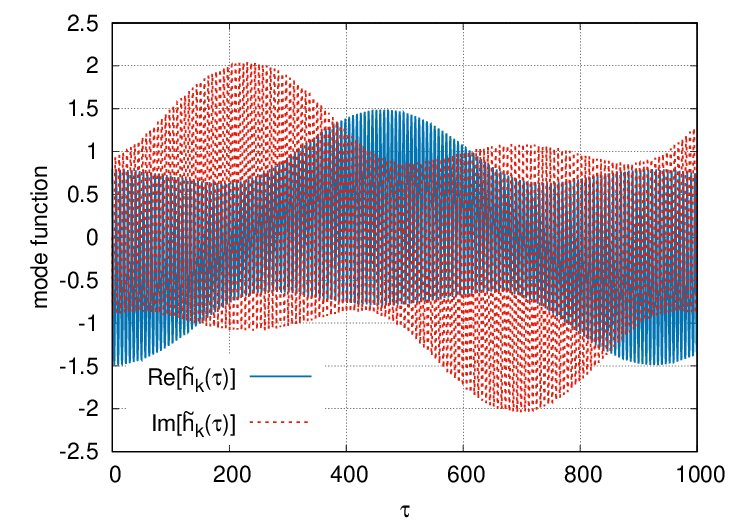} \\
		(a) The solution for parameter (i).
	\end{minipage}
	\begin{minipage}[b]{0.48\columnwidth}
		\centering
		\includegraphics[width=\linewidth]{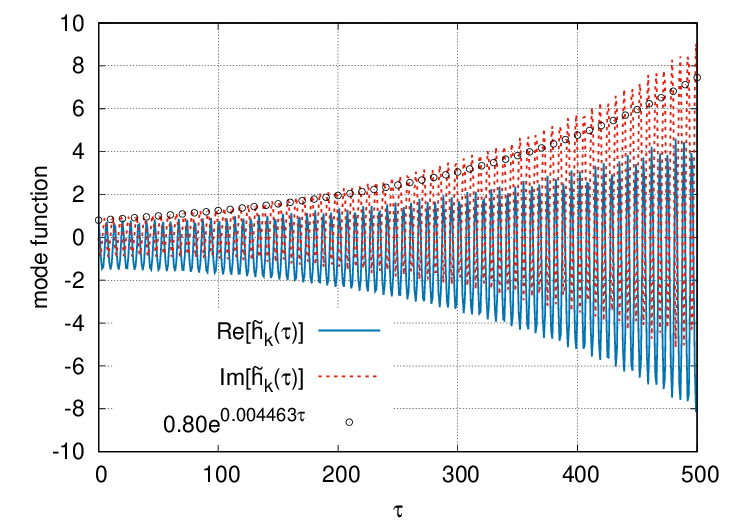} \\
	    (b) The solution for parameter (ii)
	\end{minipage}
		\caption{Two type of solutions for transverse mode function for parameter set (i) and (ii) for Z boson. The values $0.80$ and $0.004463$ are the initial value of the absolute value of the mode function and the Floquet index for the parameter (ii) \label{fig.sol.transverse.12}}
\end{figure}

The results for (i) and (ii) are shown in Figs.~\ref{fig.sol.transverse.12} and $0.80$ and $0.004463$ are the initial function value of the absolute value of the mode function and the Floquet index for the parameter (ii).
As we can see, the parameter (ii) shows the unstable behavior (exponential growth), while the parameter (i) shows the stable behavior (periodic and bound). 

\begin{figure}[htbp]
		\centering
	\begin{minipage}[b]{0.48\columnwidth}
		\centering	
	\includegraphics[width=\linewidth]{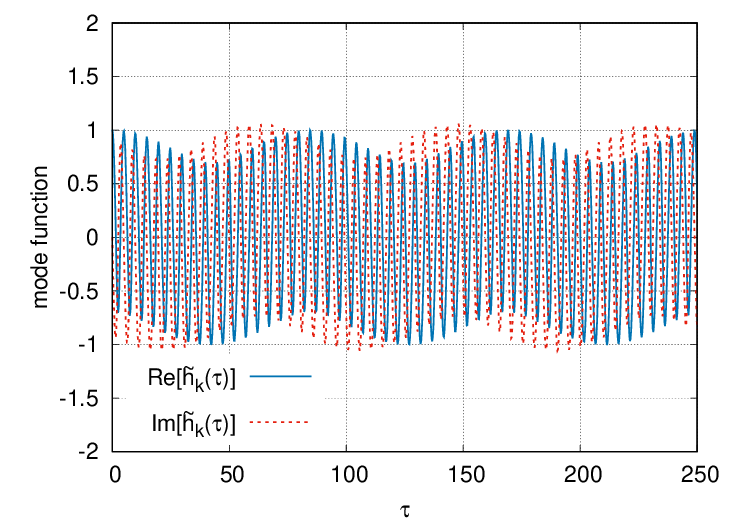} \\
	(a) The solution for parameter (iii)
	\end{minipage}
	\begin{minipage}[b]{0.48\columnwidth}
		\centering
	\includegraphics[width=\linewidth]{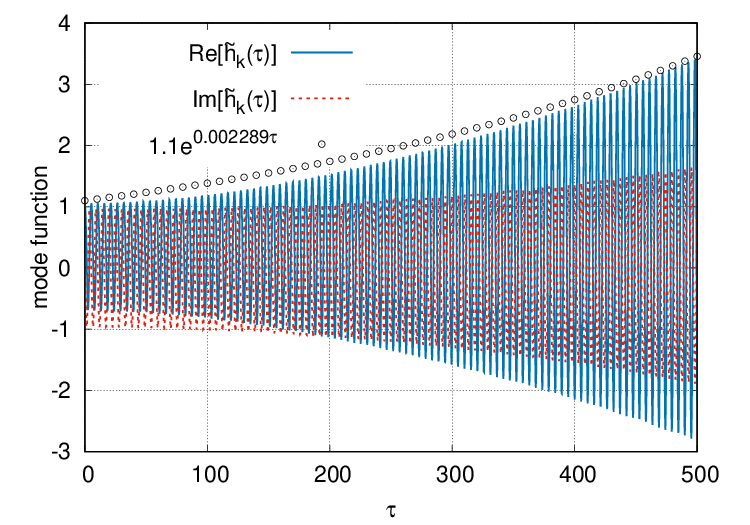}\\
(b) The solution for parameter (iv)   
	\end{minipage}
	\caption{Two type of solutions for transverse mode function for parameter set (iii) and (iv) for Z boson. The values $1.1$ and $0.002289$ are the initial value of the absolute value of the mode function and the Floquet index for the parameter (iv). \label{fig.sol.transverse.34}}
\end{figure}

Similarly, we can see the same behaviors for other parameters:
\begin{eqnarray}
	&(\mbox{iii})&~\mbox{stable~point:}~(\tilde{\phi}_0,\tilde{k}^2)=(0.4,0.8), \nn\\
	&(\mbox{iv})&~\mbox{unstable~point:}~(\tilde{\phi}_0,\tilde{k}^2)=(0.4,0.625),
\end{eqnarray}
where these parameters give $d=1.847$ for (iii) and $d=2.0002$ (Floquet index=$0.002289i$) for (iv) respectively.
The results for (iii) and (iv) are shown in Fig.~\ref{fig.sol.transverse.34} and we can see the similar behaviors with that in Fig~\ref{fig.sol.transverse.12}. We checked the Wronskian condition numerically for $u_1$ and $u_2$ in (\ref{eq.u1.u2}) and typically the deviation from unity is order of $10^{-5}$ to $10^{-4}$. 

We can obtain the longitudinal mode function $\tilde{L}_k$ with the initial conditions,
\begin{eqnarray}
 \tilde{L}_k(0) &=& \frac{\tilde{\phi}_{\mathrm{cl}}}{\tilde{\omega}_{k,L}(0)}, \hspace{1cm}
 \tilde{L}^{\prime}_k(0) 
 = -i\tilde{\phi}_{\mathrm{cl}} \tilde{\omega}_{k,L}(0)
   + \frac{\tilde{\phi}^{\prime}_{\mathrm{cl}(0)}}{\sqrt{\tilde{\omega}_{k,L}(0)}}, 
\end{eqnarray}
where the prime symbol stands for the derivative with respect to $\tau$ and $\tilde{\omega}_{k,L}(0)\equiv \sqrt{|\tilde{a}_{k,L}(0)|}$. 

\begin{figure}[htbp]
		\centering
	\begin{minipage}[b]{0.48\columnwidth}
		\centering	
		\includegraphics[width=\linewidth]{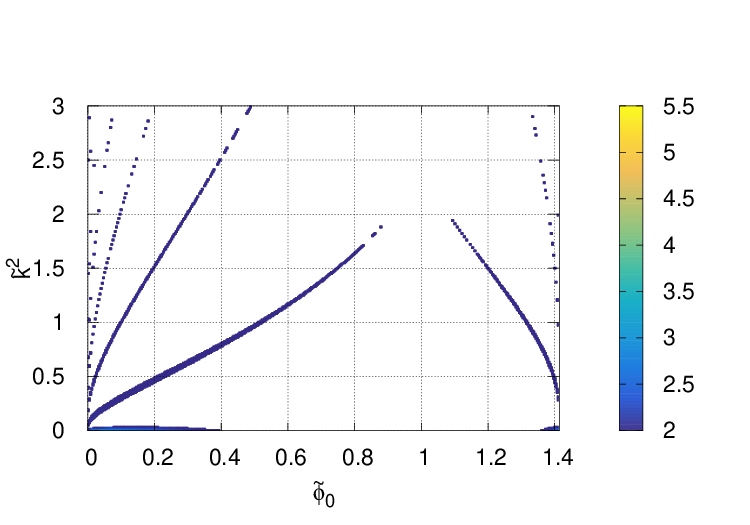}\\
	(a) Transverse mode 
	\end{minipage}
	\begin{minipage}[b]{0.48\columnwidth}
		\centering
	\includegraphics[width=\linewidth]{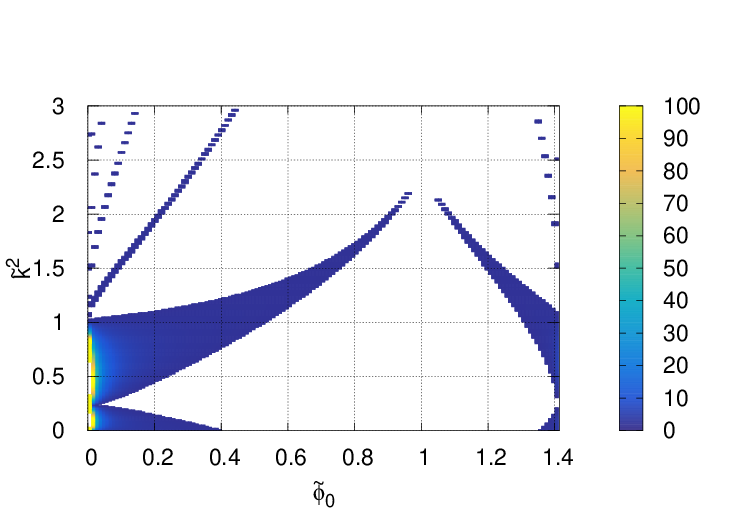}\\
(b) Longitudinal mode
	\end{minipage}
	\caption{The plot of the parameter region with $d>2$ for W case. \label{fig.disc.transverse.longitudinal.W}}
\end{figure}

Similarly, we can obtain the similar result of the instability diagrams for W boson case by setting $g_A=g_W=0.1066$. The results are shown in Fig.~\ref{fig.disc.transverse.longitudinal.W}.
 Basically the structure of the unstable line/band for W boson case is same with that for Z boson case. However, the new instability structures appear at $(\tilde{\phi}_0 < 0.4,\tilde{k}^2\approx 0)$ and at $(\tilde{\phi}_0 \approx 1.4,\tilde{k}^2\approx 0)$ for W boson case, as we discuss it in the next section.

\section{Discussion \label{Sec5}}
We discuss the physical origin of the stability/instability in the solution  $\tilde{h}_k$ in Eq.~(\ref{eq.dimless.modeeq.h}) and $\tilde{L}_{k}$ in Eq.~(\ref{eq.dimless.modeeq.L}). 

First, we focus on the instability in the transverse mode function $\tilde{h}_k$. Basically the structure of the differential equation for $\tilde{h}_k$ is directly reduced to the Hill's equation. The instability has two types in general, 1) the parametric amplification or parametric resonance which originates from the positive $\tilde{a}_{k,T}$ and 2) the spinodal instability which originates from negative $\tilde{a}_{k,T}$. The parametric resonance is often discussed in the context of cosmological particle productions \cite{Ford.2021}. 

Previous analyses include nonadiabatic vector production in a $cn$-type Standard Model Higgs background \cite{Casadio.2007} and the dynamics of dark-sector vectors \cite{Khan.2026, Dror.2010}. In contrast, the $dn$-type Higgs backgrounds considered here never cross zero, and the electroweak gauge-boson masses remain nonzero throughout the oscillation. The resulting polarization-resolved Floquet maps demonstrate that gauge-field instability does not require repeated zeros of the gauge-boson mass. They further reveal two characteristic features: broad longitudinal instability domains containing both parametric and spinodal sectors, and additional coupling-dependent low-momentum instability regions for the $W$ boson. The following comparison between $\widetilde{a}_{k,T}$ and $\widetilde{a}_{k,L}$ identifies the dynamical origin of these features.

\begin{figure}[htbp]
		\centering
	\begin{minipage}[b]{0.48\columnwidth}
		\centering	
	\includegraphics[width=\linewidth]{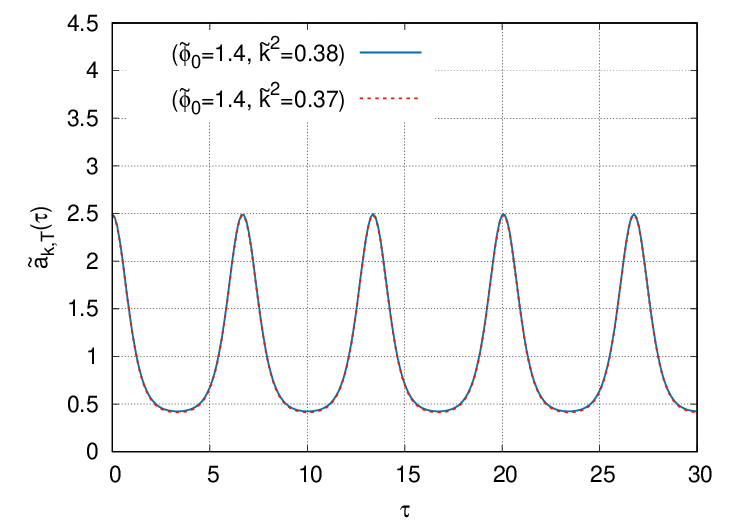}\\
		(a) The parameter set for (i) and (ii) 
	\end{minipage}
	\begin{minipage}[b]{0.48\columnwidth}
		\centering
	\includegraphics[width=\linewidth]{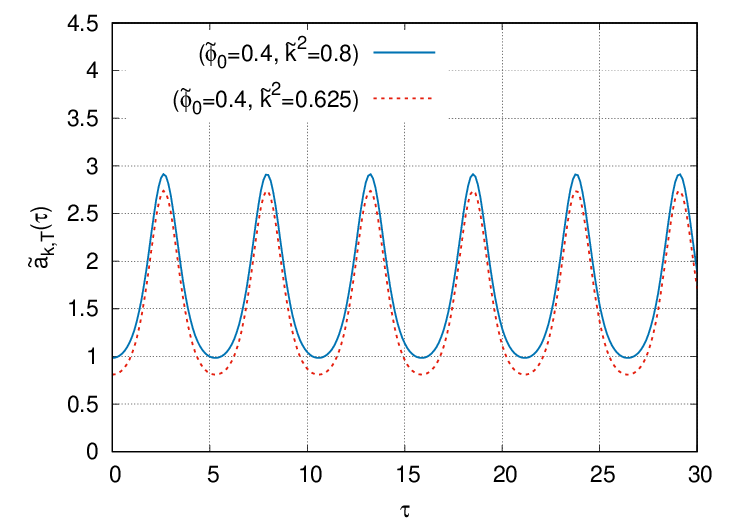}\\
	(b) The parameter set for (iii) and (iv)
	\end{minipage}
	\caption{The behavior of $\tilde{a}_{k,T}(\tau)$ for the transverse mode of Z boson. \label{fig.afuncT.transverse}}
\end{figure}

Actually the instability in $\tilde{h}_k$ is related to the parametric amplification and there is no spinodal instability. One can easily distinguish these two instabilities by checking the behavior of the periodic coefficient function $\tilde{a}_{k,T}$ of the Hill's equation in (\ref{eq.dimless.modeeq.h}). As we can see in (a) add (b) in Fig.~\ref{fig.afuncT.transverse}, the instability of (b) in Fig.~\ref{fig.sol.transverse.12} originates from the parametric amplification instability, because $\tilde{a}_{k,T}$ is obviously positive definite. However, the parametric amplification occurs for particular parameter set even if $\tilde{a}_{k,T}$ is positive. One can see such an example in classical EOM in \cite{LandauText}.
\begin{figure}[htbp]
		\centering
	\begin{minipage}[b]{0.48\columnwidth}
		\centering	
		\includegraphics[width=\linewidth]{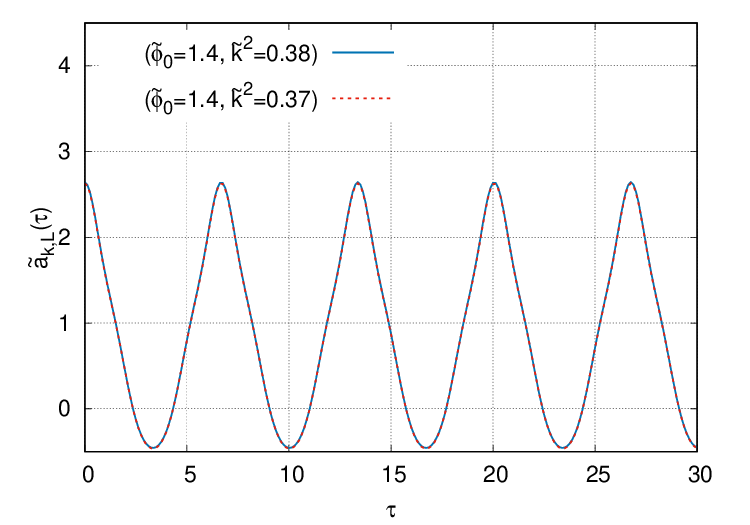}\\
		(a) The parameter set for (i) and (ii) 
	\end{minipage}
	\begin{minipage}[b]{0.48\columnwidth}
		\centering
	\includegraphics[width=\linewidth]{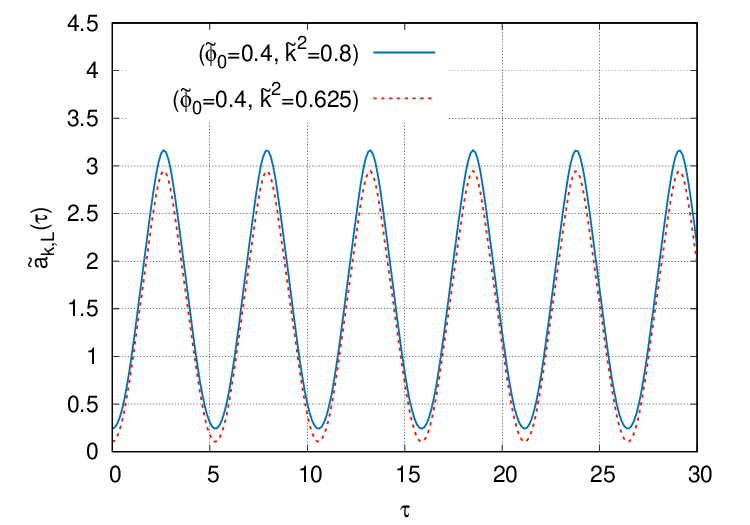}\\
	(b) The parameter set for (iii) and (iv)
	\end{minipage}
	\caption{The behavior of $\tilde{a}_{k,L}(\tau)$ for the longitudinal mode of the Z boson. \label{fig.afuncL.longitudinal}}
\end{figure} 
The difference between the nature of oscillations in (a) of Fig.~\ref{fig.sol.transverse.12} and (a) of Fig.~\ref{fig.sol.transverse.34} can be understood by the distance between the stable parameter and unstable parameter region. Typically, the stable parameter which is very close to the unstable region tends to show the long period oscillation, as we can see in (a) of Fig.~\ref{fig.sol.transverse.12}. This oscillation is the combination of long period oscillation and very short oscillation, because this parameter is close to the unstable parameter region. On the other hand, the parameter for (a) of Fig.~\ref{fig.sol.transverse.34} is relatively far from the unstable parameter region and thus the long period oscillation is not very long like that in (a) of Fig.~\ref{fig.sol.transverse.12}. 

Next, we focus on the instability in the longitudinal mode function $\tilde{L}_k$. Basically we can see the similar tendency of the difference between two oscillations in the stable parameter. However, an important difference appears in the behavior of the unstable parameter, namely, the area of the unstable parameter region for $\tilde{L}_k$ is much broader than that of for $\tilde{h}_k$. As is shown in Figs.~\ref{fig.disc.transverse.Z} and \ref{fig.disc.longitudinal.Z}, the largest unstable parameter region is like "band shape" or "horn shape" for $\tilde{L}_k$, while the corresponding largest unstable parameter region is more like line shape for $\tilde{h}_k$. 

On the longitudinal polarization case, the parameter sets (i), (ii), (iii), and (iv) reduce to unstable. By checking the behavior of $\tilde{a}_{k,L}(\tau)$, we can distinguish the origin of the instability. The instabilities in the parameter sets (i) and (ii) originate from the spinodal instability as we can see in (a) of Fig.~\ref{fig.afuncL.longitudinal}, because these parameter sets show that $\tilde{a}_{k,L}$ becomes negative in certain time region; while the instabilities in the parameter sets (iii) and (iv) originate from the parametric instability as we can see in (b) of Fig.~\ref{fig.afuncL.longitudinal}, because these parameter sets show that $\tilde{a}_{k,L}$ is positive for whole period. 

One of possible physical interpretation for the reason why the longitudinal polarization shows the stronger instability is the negative contribution to $\tilde{a}_{k,L}$ in (\ref{eq.dimless.a.L}) from EOM of the classical Higgs field described by $-\lambda(1-\tilde{\phi}^2_{\mathrm{cl}})$, except for its complex structure. This term negatively contributes to $\tilde{a}_{k,L}$ and hence the solution in (\ref{eq.dimless.modeeq.L}) tends to be more unstable than the solution in (\ref{eq.dimless.modeeq.h}), because $\tilde{a}_{k,T}$ is manifestly positive definite and hence the solutions in the most of parameter regions show the stability. Moreover, this negative tendency of the Higgs-EOM-contribution in $\tilde{a}_{k,L}$ and its instability reflect the well known fact, i.e., the creation of mass of the electroweak gauge bosons originates from the Higgs field and the longitudinal mode of the gauge field is emerged. Obviously, the structure of $-\lambda(1-\tilde{\phi}^2_{\mathrm{cl}})$ originates from the derivative term of the Higgs potential with respect to the classical Higgs field in EOM of the Higgs field, $V^{'}(\phi) = -\mu^2\phi + \lambda \phi^3 = -\lambda v^2\phi(1-\phi^2/v^2)$, where the $v\phi$ disappeared in the derivation of the dimensionless periodic coefficient function in ($\ref{eq.dimless.a.L}$). 

In short, the stability/instability diagrams in Figs.~\ref{fig.disc.transverse.Z} and \ref{fig.disc.longitudinal.Z} provide us the fundamental tools to discuss the time dynamics of the mode functions in transverse and longitudinal polarization states in the massive gauge fields like $W$ and $Z$ which couples to the nonlinear massive wave solution in the Higgs potential. Especially, this analysis of the instability in the solutions of the mode function can be carried out by the Floquet theorem in the theory of the Hill's equation. 
As a consequence, we can see the well known relation between the emergence of the mass and longitudinal polarization state of the massive electroweak boson and the Higgs potential through the expansion of the instability region in the instability diagram for the dimensionless mode function $\tilde{L}_k$ in Fig.~\ref{fig.disc.longitudinal.Z}. 
\begin{figure}[htbp]
		\centering
	\begin{minipage}[b]{0.48\columnwidth}
		\centering	
	\includegraphics[width=\linewidth]{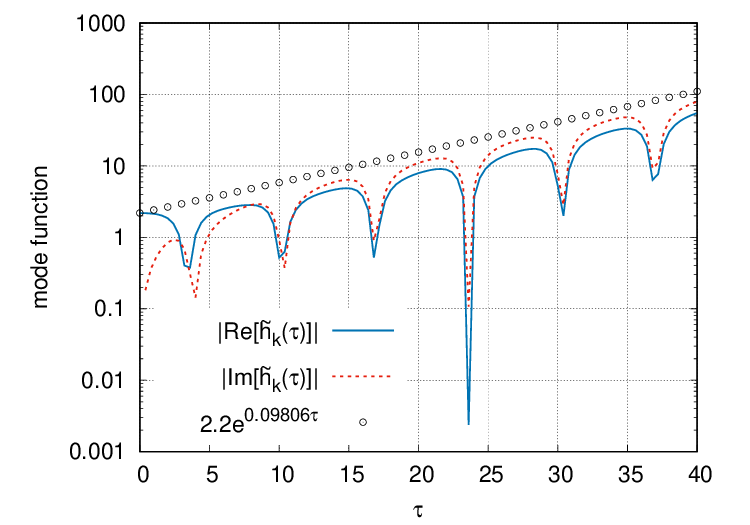} \\
	(a) Transverse mode.
	\end{minipage}
	\begin{minipage}[b]{0.48\columnwidth}
		\centering
	\includegraphics[width=\linewidth]{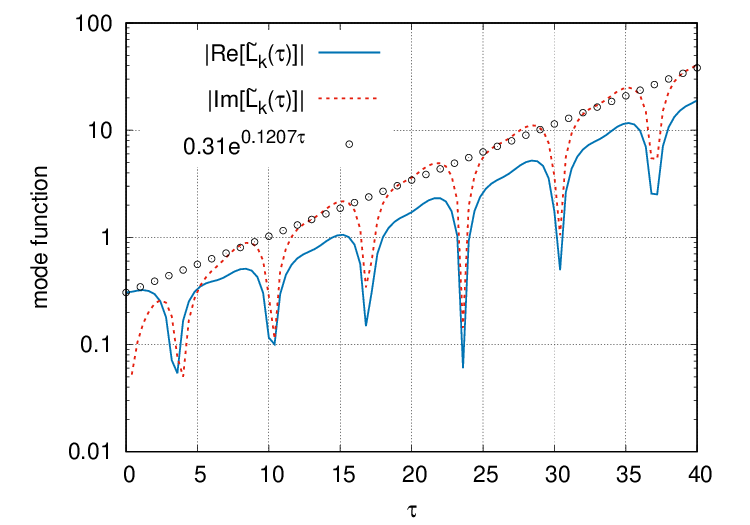} \\
	(b) Longitudinal mode.
	\end{minipage}
	\caption{The unstable solution of the mode function for W-boson with the parameter $(\tilde{\phi}_0=0.2, \tilde{k}^2=0.01)$. The values $0.31$ and $0.1207$ are the initial value of the absolute value of the mode function and the Floquet index for this parameter. \label{fig.sol.transverse.longitudinal.W}}
\end{figure}

Concerning the difference between Z boson and W boson, we can find out the important difference in the instability diagrams between Z-boson and W-boson, that is, the instability diagram for W-boson shows another instability region at the parameter $(\tilde{\phi}_0<0.4, \tilde{k}^2\simeq 0)$. We can check the behavior of the solutions in the transverse/longitudinal mode equations and the results are shown in Fig.~\ref{fig.sol.transverse.longitudinal.W}. These unstable behaviors in the mode functions for the transverse/longitudinal modes can be understood by the Floquet indexes as are shown in (a) and (b) of Fig.~\ref{fig.sol.transverse.longitudinal.W}. In this parameter region, the quantum mode in W-boson is more unstable than that in Z-boson. Hence the decays products from the W-boson with this parameter will be more important than that from the Z-boson when the classical system is described by the nonlinear massive wave solution during the electroweak phase transition.

Finally we briefly point out some possible extensions of this research. First, in this article we have not yet discussed the particle creation based on the Bogoliubov coefficient in the context of cosmological particle creation. Actually the particle creation for the transverse mode is same with that for the scalar case, because these two systems share the same structure of the differential equations for the mode equations. However, the particle creation for the longitudinal mode is much complicated, as is shown in the differential equation for $\ell_{k}$ and thus we did not discuss it in our article. Hence we must find out a suitable Bogoliubov transformation, adiabatic mode, and the Hamiltonian density of the system. Next, in order to discuss the validity of the conjecture for the emergence of a new asymptotic equilibrium state proposed in \cite{Herring.2024} in our model, we must take into account the damping effect of the field due to the evolution of the Universe \cite{Turner.1983}. As we discussed the differential equation for the longitudinal mode, its complex structure of the longitudinal mode will prevent us to analyze the detail of the new equilibrium state. Hence it will be best to consider the damping effect for the scalar case first and discuss some thermodynamic quantities like entropy density or particle density. Such an extension of the work will be more efficient to clarify the difference of the equilibrium state obtained by conventional effective potential approach and the equilibrium state proposed by the formalism in \cite{Herring.2024, Herring.2025}. 

Another important direction of the improvement of our analysis is to take into account the fermionic instability and its effect on the particle creation. As is well known, the heaviest particle in the SM is the top quark which strongly couples to the Higgs field through the Yukawa coupling. Hence, we must construct the formalism of the quantum fluctuation of the top around the nonlinear massive wave solution and particle creation effect of the top on the effective potential \cite{Herring.2024} in our model. In such a case, the analysis in Ref.~\cite{Landete.2013} will be helpful, because the authors discussed the formalism of the adiabatic mode for spin $1/2$ field in the background of evolving universe. We will discuss it in the future. 

\section{Conclusion \label{Sec6}}

We studied the stability/instability of the quantum fluctuations in the electroweak massive vector fields to find out a new aspect of the Higgs potential. We adapted the standard quantization formalism
for the transverse/longitudinal polarizations in the massive vector fields by using the method of the quantization for the fields coupling to the time-dependent external field used in cosmology. As a consequence of the research, we derived the stability/instability diagram which determines the time behavior of the quantum fluctuations in the massive vector fields, with the help of the general theory of the Hill's equation. 

The results show that the instability diagram for the longitudinal mode is broader than that for the transverse mode in electroweak gauge bosons. We argued that this expansion of the unstable region in parameter space can be understood by the coupling effect between the classical Higgs field and the quantum massive gauge field. Thus we highlighted the instability of the longitudinal mode as the uniqueness of the massive gauge fields. These results will be useful to pin down the dominant parameter region when we consider the particle creations of the quantum fields which couples to the nonlinear massive wave field during the electroweak phase transition.

The remained work is to formulate the formalism of particle creations for the longitudinal polarization in the massive gauge fields. Actually the formalism for the transverse polarization is same with that for the scalar field case, however we need subtlety for the longitudinal polarization, because of the complexity in the mode equation. Moreover, we must take into account the expansion of the Universe to discuss the particle creations which played the important role in the conjecture of the new equilibrium state proposed in \cite{Herring.2024,Herring.2025}. This is crucial to study whether such a new equilibrium state exists or not in our model of the quantum field theory coupling to the semi-stable classical solution in the Higgs potential. In particular, the effect of the evolution of the Universe will give us a hint of the adiabatic mode of the quantum scalar/vector fields when we construct the formalism of the cosmological particle creations in our model.

\section*{Acknowledgments}
Y.~K. was supported by National Science and Technology Council in Taiwan (Grant No. NSTC 115-2112-M-167-002), and by Graduate Institute of Precision Manufacturing in National Chin-Yi University of Technology. 


\end{document}